\documentstyle[aps,pre,manuscript]{revtex}

\begin{document}

\draft

\title{Nonequilibrium stochastic processes: Time dependence of 
entropy flux and entropy production}

\author{Bidhan Chandra Bag{\footnote{e-mail: pcbcb@yahoo.com}}
}

\address{Department of Chemistry, Visva-Bharati, Santiniketan 731 235, India}

\date{To appear in Physical Review E}

\maketitle

\begin{abstract}
Based on the Fokker-Planck and the entropy balance equations 
we have studied the relaxation of a dissipative dynamical system driven by
external Ornstein-Uhlenbeck noise processes
in absence and presence of nonequilibrium constraint in terms of the
thermodynamically inspired quantities like entropy flux and entropy production. 
The interplay
of nonequilibrium constraint, dissipation and noise reveals some interesting
extremal nature in the time dependence of entropy flux and entropy production.
\end{abstract}



\section{Introduction}
Understanding of the nature of nonequilibrium and equilibrium states of a dynamical
system in presence of surroundings is always an intriguing issue of
physics. Entropy is an important quantity in this regard in
thermodynamics. While in the traditional classical thermodynamics,
the specific nature of a stochastic process is irrelevant, this may play 
an important role for establishing the connection between the phase
space of a dynamical system and the related thermodynamically
inspired quantities like entropy production, flux and Onsagar coefficients
etc. Recently a number of authors \cite{pat,jt,leb,jou,gas,bb3,nic,bb4,bb2,bb1,sc1} have
explored the relationship in considerable detail.

The aim of the present paper is to enquire in this connection about the imprints of
color \cite{hr}, white and cross-correlated noise processes \cite{jz,jz3}
on time dependence of entropy, entropy production and entropy flux using 
a connection between
the information entropy and the probability distribution function of the
phase space variables for  thermodynamically open systems. 
Based on a Fokker-Planck
description of stochastic processes and the entropy balance equation
we first
consider here the relaxation of a dissipative dynamical system in presence
of the noise processes to a steady state from a given nonequilibrium
state in terms of thermodynamically inspired quantities. 
For additive white noise we compare our results in the equilibrium
state with the standard results for the closed systems. 
We also enquire how the system relaxes if the system is thrown away from the
aforesaid steady state by a nonequilibrium constraint to understand
how the entropy flux and the entropy production pass through minima with time
in the later case and how the two
relaxation processes for different noise properties differ. 

The outline of the paper is as follows: In Sec. II we 
calculate the entropy flux and the entropy production for a simple dissipative
dynamical system in the nonequilibrium state 
for different noise
processes. The paper is concluded in the Sec.III.

\section{The Fokker-Planck description, time dependence of 
entropy flux and production of noise-driven dynamical systems}

\subsection{Relaxation of the noise-driven dynamical system to the steady state}

\subsubsection{Ornstein-Uhlenbeck noise process}

We consider the dynamics of a dissipative dynamical system driven by the
external Ornstein-Uhlenbeck noise process in the phase space. The relevant
Langevin equation of motion can be written as

\begin{equation}
\dot{X} =  -\gamma X + \eta
\end{equation}

\noindent
where $\gamma$ is the damping constant. The term $\eta$ in Eq.(1) is 
the external Ornstein-Uhlenbeck noise whose two time correlation is given by

\begin{equation}
\langle \eta (t) \eta (t') \rangle = \frac{D}{\tau} 
\exp \left ( -\frac{|t-t'|}{\tau} \right ) \; \; \; .
\end{equation}

\noindent
$D$ is the noise strength and $\tau$ corresponds to the correlation time
of colour noise process. The time evolution of $\eta$ can be conveniently
expressed in terms of the Gaussian white noise process $\zeta(t)$ as

\begin{equation}
\dot{\eta} = -\frac{\eta}{\tau} +\frac{\sqrt D}{\tau} \zeta
\end{equation}

\begin{eqnarray*}
\langle \zeta(t) \zeta(t') \rangle = 2 \epsilon \delta(t-t')
\end{eqnarray*}

\noindent
and

\begin{eqnarray*}
\langle \zeta \rangle = 0   \; \ ;,
\end{eqnarray*}

\noindent
here the parameter $\epsilon$ is used to identify the noise strength. 

Now treating $\eta$ as a phase space  variable on the same footing as $X$
we can write Fokker-Planck in the extended phase space \cite{hr} as

\begin{eqnarray}
\frac{\partial \rho(X_1, X_2, t)}{\partial t} & = & \gamma \frac{\partial X_1 \rho}{\partial X_1}
-X_2 \frac{\partial \rho}{\partial X_1}
+\frac{1}{\tau} \frac{\partial X_2\rho}{\partial X_2} + \epsilon \frac{D}{\tau^2}
\frac{\partial^2 \rho}{\partial X_2^2} 
\end{eqnarray}

\noindent
where $X_1, X_2$ refer to $X$ and $\eta$ in Eq.(1) and $\rho(X_1, X_2, t)$ is
the extended phase space probability distribution function.

Now making use of the following transformation

\begin{equation}
U=a X_1+ X_2  \; \; \;,
\end{equation}

The Fokker-Planck Eq.(4) can be written as

\begin{equation}
\frac{\partial \rho(U,t)}{\partial t} = -\frac{\partial F \rho}{\partial U}  
+\epsilon  D_s\frac{\partial^2 \rho}{\partial U^2}  \; \; ,
\end{equation}

\noindent
where 
\begin{equation}
F =-\lambda U  \; \; \; ,
\end{equation}

\begin{equation}
\lambda U = \gamma a X_1 -a X_2 + \frac{X_2}{\tau} \; \; , \; \;
\end{equation}

\noindent
and

\begin{equation}
D_s = \frac{D}{\tau^2} \; \; .
\end{equation}

Here $a$ and $\lambda$ are constants to be determined. 
Using Eq.(5) in
Eq. (8) and comparing the coefficients of $X_1$ and $X_2$  we find

\begin{equation}
\lambda  = \gamma  \; \; {\rm and} \; \; 
a =  \frac{1-\gamma \tau}{\tau} \; \; .
\end{equation}

We are now in a position to define entropy flux and entropy production
using Eq.(6). In the microscopic picture the Shannon form of the 
entropy is connected to the continuous probability distribution $\rho$ as

\begin{equation}
S = - \int \rho(U, t) \ln \rho(U, t) du
\end{equation}

The time evolution equation for entropy then can be written as

\begin{equation}
\frac{dS}{dt} = -\int dU \left[-\frac{\partial F\rho}{\partial U} +\epsilon D_s
\frac{\partial^2 \rho}{\partial U^2} \right] \ln \rho
\end{equation}

Putting the usual boundary conditions into the result of partial integration of
the right hand side of the above equation (12), one obtains the following
form of information entropy balance

\begin{equation}
\frac{dS}{dt} = \int \rho \frac{\partial F}{\partial U} dU 
+\epsilon D_s \int \frac{1}{\rho}\left(\frac{\partial \rho}{\partial U} \right)^2 dU 
\end{equation}

Eq.(13) implies that the first term has no definite sign while the second term
is positive definitely since $D_s$ is always positive. Then one can identify
the first and the second terms as entropy flux $(\dot{S}_F)$ and entropy production
$(\dot{S}_P)$, respectively.

\begin{equation}
\dot{S}_F =\int \rho \frac{\partial F}{\partial U} dU
\end{equation}

\begin{equation}
\dot{S}_P =\epsilon D_s \int \frac{1}{\rho} \left(\frac{\partial \rho}{\partial U} \right)^2 dU 
\end{equation}

To find the explicit time dependence of these quantities
we then search for the Green's function or conditional probability
solution for the system at $U$ at time  $t$ 
for the given initial condition 

\begin{equation}
\rho(U, t=0) =  \frac{\epsilon_1}{\pi}
\exp[-\epsilon_1 (U-U')^2]
\end{equation}

We now look for a solution of the Eq.(6) of the form

\begin{equation}
\rho(U, t|U', 0) = \exp[G(t)]
\end{equation}

\noindent
where
\begin{equation}
G(t) = -\frac{1}{\sigma(t)} (U-\beta(t))^2 +\ln \nu(t)
\end{equation}

We will see that by suitable choice of $\beta(t), \sigma(t), \nu(t)$
one can solve Eq.(6) subject to the initial condition

\begin{equation}
\rho(U, 0|U', 0) =  \frac{\epsilon_1}{\pi}
\exp[-\epsilon_1 (U-U')^2] \; \;.
\end{equation}

Comparing Eq.(19)  with (17) and $G(0)$ we have

\begin{equation}
\sigma(0)=\frac{1}{\epsilon_1}, \; \; \; \beta(0)=U', \; \; \;  \nu(0)=\frac{\epsilon_1}{\pi} \; \;.
\end{equation}

If we put (17) in (6) and equate the coefficients of equal powers of $U$
we obtain after some algebra the following set of equations

\begin{equation}
\dot{\sigma(t)} = -2\gamma \sigma(t) +4\epsilon D_s
\end{equation}

\begin{equation}
\dot{\beta(t)} =  -\gamma\beta(t) 
\end{equation}

\begin{equation}
\frac{1}{\nu(t)}\dot{\nu(t)} = -\frac{1}{2\sigma(t)}\dot{\sigma(t)}
\end{equation}

The relevant solutions of $\sigma(t)$ and $\beta(t)$ for the present problem which
satisfy the initial conditions above are given by

\begin{equation}
\sigma(t) = \frac{2\epsilon D_s}{\gamma}(1-\exp(-2\gamma t))+ \sigma(0) \exp(-2\gamma t)
\end{equation}

\noindent
and

\begin{equation}
\beta(t) = \beta(0) \exp(-\gamma t)
\end{equation}

Now making use of Eqs. (17), (24) and (25) in Eqs.(14) and (15) 
we finally obtain the explicit
time dependence of the entropy flux and the entropy production as

\begin{equation}
\dot{S}_F =-\gamma
\end{equation}

\noindent
and

\begin{equation}
\dot{S}_P =\frac{2\epsilon D}{\tau^2[\frac{2\epsilon D}{\gamma \tau^2}+(\sigma(0)-
\frac{2\epsilon D}{\gamma \tau^2}) \exp(-2\gamma t)]}
\end{equation}

\noindent
respectively, where we have used $D_s =\frac{D}{\tau^2}$. Thus entropy flux is negative and is
independent  of time, noise strength and correlation time. But
entropy production decreases monotonically almost exponentially with
time for a given set of $D, \tau$ and $\gamma$ as shown in Fig.1 and 
finally reaches to the limiting value $\gamma$ at the long time
satisfying ($\dot{S}_F =-\dot{S}_P$)

\begin{equation}
\frac{dS}{dt} =\dot{S}_F +\dot{S}_P=0  \; \; \; .
\end{equation}

We now examine the connection between the thermodynamic entropy production 
and the phase-space collapse of the systems in nonequilibrium stationary states.
In this state $\frac{dS}{dt}=0$ and we have from Eqs. 13, 14 and 15 (for
details see Ref.[7])

\begin{equation}
\dot{S_P} =-\dot{S_F} = -\int \rho \frac{\partial F}{\partial U} dU =-\overline{div F^{\infty}}
=-\sigma'+O(\epsilon)>0
\end{equation}

\noindent
in the limit $\epsilon <<1$. Here $\sigma'$ is the lyapunov exponent of the
one dimensional deterministic system. Thus information entropy as defined
by Eq.(15) is equal to the negative of Lyapunov exponent or equivalently to
the rate of phase space volume contraction plus a correction term 
vanishing as the
noise strength goes to zero \cite{nic1,nic2}. The results in Eq.(29) is very much
interesting, since it would seem at first sight from Eq.(15) that $\dot{S_P}$
should tend to zero as $\epsilon\rightarrow 0$. The fact is 
that it nevertheless
gives a finite contribution in this limit which reflects the nonanalytic dependence
of the probability density on $\epsilon$ [7].

\subsubsection{Cross-correlated noise process}

We now consider another case where a simple dissipative system is driven
by both additive and multiplicative white Gaussian noises

\begin{equation}
\dot{X} =-\gamma X -\zeta_1 X +\eta_1
\end{equation}

The correlation between the noise processes are given by

\begin{eqnarray*}
\langle\zeta_1(t)\zeta_1(t')\rangle =2 \epsilon D'\delta(t-t')
\end{eqnarray*}

\begin{eqnarray*}
\langle\eta_1(t)\eta_1(t')\rangle =2 \epsilon \alpha \delta(t-t')
\end{eqnarray*}

\begin{equation}
\langle\zeta_1(t)\eta_1(t')\rangle = \langle\zeta_1(t')\eta_1(t)\rangle = 
2 \lambda_1 \epsilon \sqrt{D'\alpha}\delta(t-t'), \; \; \; \; 0\leq \lambda_1 \leq 1
\end{equation}

\noindent
where $\lambda_1$ denotes the cross-correlation of the two noise processes.
The Fokker-Planck equation for the Langevin Eq.(30) can be written as
(for details see \cite{bb4}) 

\begin{equation}
\frac{\partial \rho}{\partial t} = -\frac{\partial F \rho}{\partial X}
+ \epsilon D_1 \frac{\partial^2 \rho}{\partial X^2}
\end{equation}

\noindent
where the drift term is

\begin{equation}
F = -\Gamma X+l
\end{equation}

\noindent
and

\begin{equation}
D_1 =[\alpha \gamma^2 +(2-\nu)\epsilon D'\alpha\{(2-\nu)\epsilon D'
+2\gamma-2\gamma {\lambda_1}^2
-{\lambda_1}^2(2-\nu)\epsilon D'\}]/{\Gamma^2}
\end{equation}

\noindent
with

\begin{equation}
\Gamma = \gamma+2\epsilon D'-\nu \; \; \;, l=(2-\nu)\lambda_1\epsilon \sqrt{D'\alpha}
\end{equation}

In Eqs. (34) and (35) $\nu =1$ stands for the Stratonovich and $\nu =0$ for
the Ito convention.

The Fokker-Planck equation (32) is very similar to 
Eq.(6). Following the earlier method 
the time dependence of entropy flux and entropy production
for the cross-correlated noise-driven process is

\begin{equation}
\dot{S}_F =-\Gamma
\end{equation}

\noindent
\begin{equation}
\dot{S}_P =\frac{2D_1}{\sigma_1(t)}
\end{equation}

\noindent
where
\begin{equation}
\sigma_1(t) = \frac{2\epsilon D_1}{\Gamma}+(\sigma_1 (0) -\frac{2\epsilon D_1}{\Gamma})
\exp(-2\Gamma t)
\end{equation}

Here $\sigma_1(0)$ has the same significance as in Eq.(24). Thus entropy
flux for the cross-correlated noise process is time independent but its value
not only depends on dissipation constant $\gamma$ 
as in the  previous case 
but also on the strength of multiplicative noise($D'$).
The time dependence of entropy production is qualitatively same as in the Fig.1
but the relaxation time is different since $\Gamma$ contains both $\gamma$ and
$D'$. In the long time limit Eqs.(36) and (37) satisfy Eq.(28). Since
Eqs. (6) and (32) are formally same, the connection betwen the thermodynamic  
entropy production and the phase-space collapse of systems in nonequilibrium
stationary states for the correlated noise driven system should be similar to
Eq.(29). Using $D'=0, \lambda_1 =0, \nu=0$ and $\alpha=\gamma K T$ in Eq.(37)
($K$ and $T$ are Boltzmann constant and temperature, respectively) one can
obtain the time dependence of entropy flux and production for
thermodynamically closed system \cite{lw} in the Markovian limit.

\subsection{Relaxation of small external force-driven steady state to the
new steady state}

\subsubsection{The Ornstein-Uhlenbeck noise process}

We shall now examine the time dependence of entropy flux and production
during the relaxation of steady state to a new steady state for the system
driven by an weak external force. 
To this end we consider the constant drift $f_e$ in Eq.(1)
due to external force so that the total drift in Eq.(6) now becomes

\begin{equation}
F = F_0(U)+hF_1
\end{equation}

\noindent
where $F_0=-\lambda U, F_1=a f_e$ and h is smallness parameter. When 
$h=0$, $\rho= \rho_s$, $\rho_s$ is the steady state solution of the
Eq.(6). The deviation of $\rho$ from $\rho_s$ in presence of nonzero small
$h$ can be explicitly taken into account once we make use of the identity
for the diffusion term in Eq.(6)

\begin{equation}
\frac{\partial^2 \rho}{\partial U^2} = \frac{\partial}{\partial U}
\left[\rho \frac{\partial \ln \rho_s}{\partial U}\right]
+\frac{\partial}{\partial U}
\left[\rho_s \frac{\partial}{\partial U}\frac{\rho}{\rho_s}\right]
\end{equation}

Now we are in a position to establish a connection between the entropy 
production of irreversible thermodynamics and the relevant quantities 
of the underlying
dynamics in phase space for the present model following Ref.[7]. The explicit
calculation using Eq.(40) shows that the information entropy balance Eq.(12) 
now yields

\begin{eqnarray}
\frac{dS}{dt} & = & 
-\int dU \ln \rho \left[-\frac{\partial (F \rho)}{\partial U}
+ \epsilon D_s \frac{\partial}{\partial U}
\left ( 
\frac{\rho \partial \ln \rho_s}{\partial U} \right )
\right] \nonumber \\
& & - \epsilon D_s \int dU \ln \rho_s \frac{\partial}{\partial U} 
\left ( \rho_s \frac{\partial}{\partial U} \frac{\rho}{\rho_s} \right )
+ \epsilon D_s \int dU \rho \left ( \frac{\partial}{\partial U}
\ln \frac{\rho}{\rho_s} \right )^2
\end{eqnarray}

It is noted that the first , the second and the third integrals in Eq.(41) are of zeroth,
first and second order, respectively, with respect to the deviation from
steady state. Doing partial integrations in Eq.(41) we obtain

\begin{equation}
\frac{dS}{dt} =\overline{div F^t}
+ \epsilon D_s \int dU \rho 
\left [ - \left ( \frac{\partial \ln \rho_s}{\partial U} \right )^2
+ 2 \frac{\partial \ln \rho}{\partial U} 
\frac{\partial \ln \rho_s}{\partial U} \right ] 
+ \epsilon D_s \int dU \rho \left ( \frac{\partial}{\partial U}
\ln \frac{\rho}{\rho_s} \right )^2
\end{equation}

Such a new decomposition of the rate of change of information entropy now
exhibits a part $\dot{\Delta S_P}$

\begin{equation}
\dot{\Delta S_P} =\epsilon D_s \int dU \rho (\frac{\partial}{\partial U}
\ln \frac{\rho}{\rho_s})^2\ge0
\end{equation}

\noindent
which is both positive definite and of second order in the deviation from the
steady state, thereby fulfilling the principal condition required on entropy production.
On the otherhand, the first term on the right-hand side of Eq.(42),
$\overline{div F^t}$, has no definite sign and contains, in principle, contributions
of all orders in the deviation from steady state. In the stationary state, 
$\frac{dS}{dt}=0$, and the contribution of this term and of the 
second one in Eq.(42) must cancel that of $\dot{\Delta S_P}$. The role of this latter
term in this balance is, then, to remove the contributions of all but second
orders in the deviation from steady state contained in $\overline{div F^t}$.

We may therefore write, in the new steady state

\begin{equation}
\dot{\Delta S_P} = - \overline{div F^\infty} -
(\rm{terms\; of \; 0th \; and \; 1st \; order \; in}\; h) .
\end{equation}

\noindent
So by virtue of Eq.(29) we have

\begin{equation}
\dot{\Delta S_P} =-\sigma' - (\rm{terms\; of \; 0th \; and \; 1st \; order \; in}\; h) .
\end{equation}

This establishes a connection between the irreversible thermodynamics on the one
hand, and phase space dynamics on the other in the case when the dynamical
system is externally driven by deterministic small term.

We now return to Eq.(6) and consider the dynamics in presence of an additional
force $h F_1$ (Eq.34)

\begin{equation}
\frac{\partial \rho}{\partial t} = -\frac{\partial \phi \rho}{\partial U}
-h\frac{\partial F_1 \rho}{\partial U} + 
D_s \frac{\partial}{\partial U} 
\left(\rho_s \frac{\partial}{\partial U}\frac{\rho}{\rho_s}\right)
\end{equation}

\noindent
where $\phi$ is defined as

\begin{equation}
\phi = F_0 -D_s \frac{\partial \ln \rho_s}{\partial U} \; \;.
\end{equation}

\noindent
Here we have used $\epsilon=1$ for the rest of the calculation. 

The steady state solution of Eq.(6) is

\begin{equation}
\rho_s = N \exp[-\frac{\lambda U^2}{2D_s} ]
\end{equation}

\noindent
where $N$ is the normalization constant.

Using Eq.(48) in (47) we have

\begin{equation}
\phi \rho_s = 0
\end{equation}

To consider the entropy flux and the entropy production in the nonequilibrium state in
presence of external forcing we use Eq.(46) in the time evolution equation
of entropy(11). Following Ref.\cite{nic} we finally identify entropy flux ($\dot{\Delta S_F}$)
and entropy production ($\dot{\Delta S_P}$) as

\begin{equation}
\dot{\Delta S_F} = -\frac{d}{dt} \int \rho \frac{d \ln \rho_s}{d U} dU
+\int \frac{dF_1}{dU} \delta \rho dU +\int dU (F_1 \frac{d \ln \rho_s}{dU}) \delta \rho
\end{equation}

\noindent
and
\begin{equation}
\dot{\Delta S_P} = D_s \int dU \rho \left(\frac{d}{dU}\ln \frac{\rho}{\rho_s}\right)^2 \; \; \;.
\end{equation}

Here we have used $\delta \rho = \rho-\rho_s$ and $h=1$.

In the next step we solve Eq.(46) as before to find the explicit time
dependence of $\dot{\Delta S_F}$ and $\dot {\Delta S_P}$. The time dependent
solution of Eq.(46) is given by

\begin{equation}
\rho = N_1 \exp[-\frac{(U-\beta_h(t))^2}{\sigma(t)}]
\end{equation}

\noindent
where $N_1$ is the normalization constant and $\sigma(t)$ is obtained from
Eq.(24). The expression for $\beta_h(t)$ is given by

\begin{equation}
\beta_h(t) = \frac{F_1}{\lambda} + (\beta_h(0)-\frac{F_1}{\lambda}) \exp[-\lambda t]
\end{equation}

Now using Eqs.(48) and (52) in both (50) and (51) we have

\begin{equation}
\dot{\Delta S_F} = \frac{\lambda}{2D_s} \left[2D_s -\lambda \sigma(t) +
2 \beta_h (-\beta_h\lambda+F_1)\right]-\frac{\lambda}{D_s}F_1 \beta_h
\end{equation}

\noindent
and

\begin{equation}
\dot{\Delta S_P} = D_s\left[(\frac{\lambda}{D_s}-\frac{2}{\sigma(t)})  
\{(\frac{\lambda}{D_s}-\frac{2}{\sigma(t)}) ({\beta_h}^2 +\frac{\sigma(t)}{2}) 
+4 \frac{{\beta_h}^2}{\sigma(t)}\}+4 \left(\frac{\beta_h}{\sigma}\right)^2\right]
\end{equation}

\noindent
where $\lambda, D_s, \sigma(t), \beta_h(t)$ and $F_1$ are given by the 
Eqs. (10), (9), (24), (53) and (39) respectively. The time dependence
of $\dot{\Delta S_P}$ is shown in Fig.2 for different values of $\tau$ for 
a given set of values of other parameters. It is interesting to note that for
$\gamma \tau \neq 1$ the entropy production first decreases with time and then
passes through the minima and finally reaches to the following steady value \cite{bb4} which is
shown by solid curve of Fig.2.

\begin{equation}
\dot{\Delta S_P} = \frac{(1-\gamma \tau)^2 f_e^2}{D} = -\dot{\Delta S_F}
\end{equation}

This observation can be explained by simplifying Eq.(55) in the limit 
$\sigma(0) \rightarrow 0$ and $\beta_h(t) \rightarrow 0$ as

\begin{equation}
\dot{\Delta S_P} = \frac{1}{D (1-\exp(-2 \gamma t)} \left[(1-\gamma\tau)^2 f_e^2
(1-2 e^{-\gamma t} +2 e^{-3\gamma t}-e^{-4\gamma t})+\gamma D e^{-4\gamma t} \right]
\end{equation}

In Eq.(57) first term in the numerator which vanishes as 
$t \rightarrow 0$ implies
that the external force increases entropy production while the second
term corresponds the decrease of entropy production with time due to
dissipative action. Because of these two opposite effects a system thrown away
from a steady state by a small external force relaxes to a
new steady state passing through a minima in entropy production with time
for the case $\gamma \tau \neq 1$. For $\gamma \tau =1$ entropy
production decreases monotonically since the effective external force becomes zero
under this condition. Similarly entropy flux also 
show extremum properties for $\gamma \tau \neq 1$ case which is shown in
the solid curve of Fig.3. Dotted curve of this figure corresponds to the
time dependence of entropy flux for $\gamma \tau =1$. Another interesting point which
should be noted here is
that $\frac{dS}{dt}$ and $\dot{\Delta S_P}$ or $\dot{\Delta S_P}$
reach their equilibrium values at different times (the plot of $\frac{dS}{dt}$
vs t is shown in the inset of Fig.2). Thus Fig.2 implies that before the
true stationary state is reached the system may show $\frac{dS}{dt}=0$.

In the Markovian limit $\tau \rightarrow 0$ so that Eq.(57) reduces to

\begin{equation}
\dot{\Delta S_P} = \frac{1}{D (1-\exp(-2 \gamma t))} \left[ f_e^2
(1-2 e^{-\gamma t} +2 e^{-3\gamma t}-e^{-4\gamma t})+\gamma D e^{-4\gamma t} \right]
\end{equation}

The above equation implies that even for white noise entropy production
passes through the minima with time for both thermodynamically open and
closed ($D=\gamma K T$) systems \cite{lw}. As $t \rightarrow \infty$ 
the Eq.(58) reduces to

\begin{equation}
\dot{\Delta S_P} = \frac{f_e^2}{D} 
\end{equation}

For $D=\gamma K T$ the above equation reduces to the standard result for
entropy production of irreversible processes for a Brownian oscillator.

Eq.(57) further implies that for $t>0$
the entropy production  $\dot{\Delta S_P}$ passes through minimum
at $\gamma \tau =1$ 
which is shown in Fig.4. The variation of  $\dot{\Delta S_F}$ with $\tau$ in
Eq.(54) shows the maximum as evident in Fig.5. 
These extremal behaviour is not observed for $h=0$.

Now to show the effect of $\gamma$ on the interplay between
$\gamma$ and $\tau$ we plot both $\dot{\Delta S_P}$ vs $\gamma$
and $\dot{\Delta S_F}$ vs $\gamma$ using Eq.(55) and (54). Both the figures 
show extremum properties but Eqs. (26) and (27) do not exhibit 
such kind of variation.
It is thus apparent that in presence of the nonequilibrium
constraint the properties of noise processes as well as the dynamical
characteristic of the system are important for both entropy
flux and production.

\subsubsection{Cross-correlated noise driven process}

We now turn again to the cross-correlated noise driven process to study
the time dependence of entropy flux and entropy production due to additional
weak forcing on the stationary system. To this end we add a constant of
force $f_e$  in the Eq.(30)

\begin{equation}
\dot{X} =-\gamma X -\zeta_1 X +\eta_1 +h f_e
\end{equation}

The Fokker-Planck equation corresponding to Eq.(60) can be written as

\begin{equation}
\frac{\partial \rho}{\partial t} = -\frac{\partial \phi_1 \rho}{\partial X}
-h\frac{\partial f_e \rho}{\partial X} + 
D_1 \frac{\partial}{\partial X} 
\left(\rho_s \frac{\partial}{\partial X}\frac{\rho}{\rho_s}\right)
\end{equation}

\noindent
where 

\begin{equation}
\phi_1 = F -D_1 \frac{\partial \ln \rho_s}{\partial X}
\end{equation}

\noindent
$F$ is given by Eq.(33) and $\rho_s$ is stationery solution of Eq.(32).
Using $\rho_s$ in $\phi_1 \rho_s$ again we have

\begin{equation}
\phi_1 \rho_s =0
\end{equation}

Since Eq. (61) is very much similar to the Eq.(32), the time dependence of
entropy flux and entropy production can be derived as before to obtain

\begin{equation}
\dot{\Delta S_F} = \frac{\Gamma}{2D_1} \left[ 2D_1 -\Gamma \sigma_1(t) +
2 (\beta_h'-\frac{l}{\Gamma}) (-\beta_h'\Gamma+l+f_e)\right]
+\frac{lf_e}{D_1}-\frac{\Gamma}{D_1}f_e \beta_h'
\end{equation}

\noindent
and

\begin{equation}
\dot{\Delta S_P} = D_1\left[(\frac{\Gamma}{D_1}-\frac{2}{\sigma_1(t)})  
\{(\frac{\Gamma}{D_1}-\frac{2}{\sigma_1(t)}) ({\beta_h'}^2 +\frac{\sigma_1(t)}{2}) 
+2 (\frac{2\beta_h'}{\sigma_1(t)}-\frac{l}{D_1}) \beta_h'\}
+ (2\frac{\beta_h'}{\sigma_1}-\frac{l}{D_1})^2\right]
\end{equation}

\noindent
where

\begin{equation}
\beta_h'(t) = (\beta_h'(0) -\frac{l+f_e}{\Gamma}) \exp(-\Gamma t) + 
\frac{l+f_e}{\Gamma}
\end{equation}

Eqs.(65) and(64) also show extremal properties as shown by solid curves in Figs.
2 and 3. The variation of both $\dot{\Delta S_F}$ and $\dot{\Delta S_P}$
with noise correlation strength $\lambda_1$ in Eqs. (64) and (65) is
shown in Figs. (9) and (8) respectively at $t=0.5$. Although both the
figures show  extremal behaviour in the nonequilibrum state but at the stationary state
$\dot{\Delta S_P}$ increases and $\dot{\Delta S_F}$ decreases monotonically.
Thus the interplay between $\gamma$, noise strength and cross-correlation strength 
in the nonequilibrium state is different from that in the stationary state.
Before leaving this section we mention here that our calculated entropy flux
and entropy production are exact since the models considered here are
linear and are exactly solvable by Greens' function of 
Gaussian form.

\section{Conclusions}

In this paper we have explored the interplay between dissipative 
characteristics
of the dynamics and noise properties in presence and absence of 
nonequilibrium constraint 
in the nonequilibrium state as well as in the stationary 
state in terms of entropy flux and entropy production. Both the
entropy production and the entropy flux  show extremal properties with
time for color noise processes when the product of correlation time
and dissipation constant is not equal to one in presence of a nonequilibrium
constraint. The white and the cross-correlated noise driven processes also
mimic this extremal nature. This is due to a competition
between the nonequilibrium constraint and the dissipative action. 
The maxima and minima are also found in the variation of both 
$\dot{\Delta S_F}$ and $\dot{\Delta S_P}$ with correlation time and
dissipation constant for the color noise driven processes in the 
nonstationary and the stationary states but this feature can be found in the
variation of $\dot{\Delta S_F}$ and $\dot{\Delta S_P}$ as a function of 
correlation strength $\lambda_1$ only in the nonstationary state. Since 
white, color or cross-correlated noise driven
processes concern many situations in biology, physics and chemistry we hope that our
present observation will be useful for understanding the close connection
between irreversible thermodynamics and dynamical system in many related issues.

\acknowledgments
The author expresses his deep sense of gratitude to Prof. D S Ray 
for his kind attention throughout its progress.

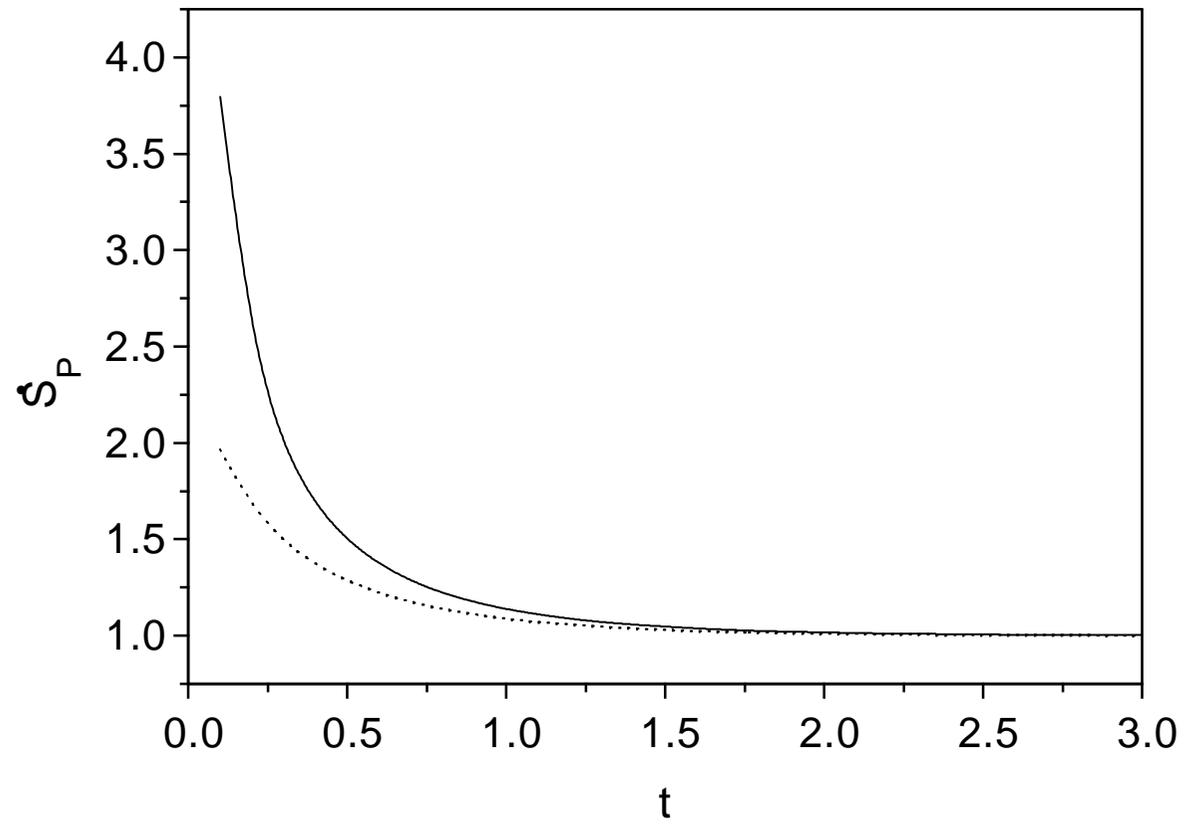
\begin{figure}
\caption{
Plot of entropy production ($\dot{S_P}$) vs time using Eq.(27) for
$\sigma(0)=0.1$, $D=0.5$, $\gamma =1.0$. Solid and dotted curves are
for $\tau=2$ and 1 respectively (Units are arbitrary).
}
\end{figure}

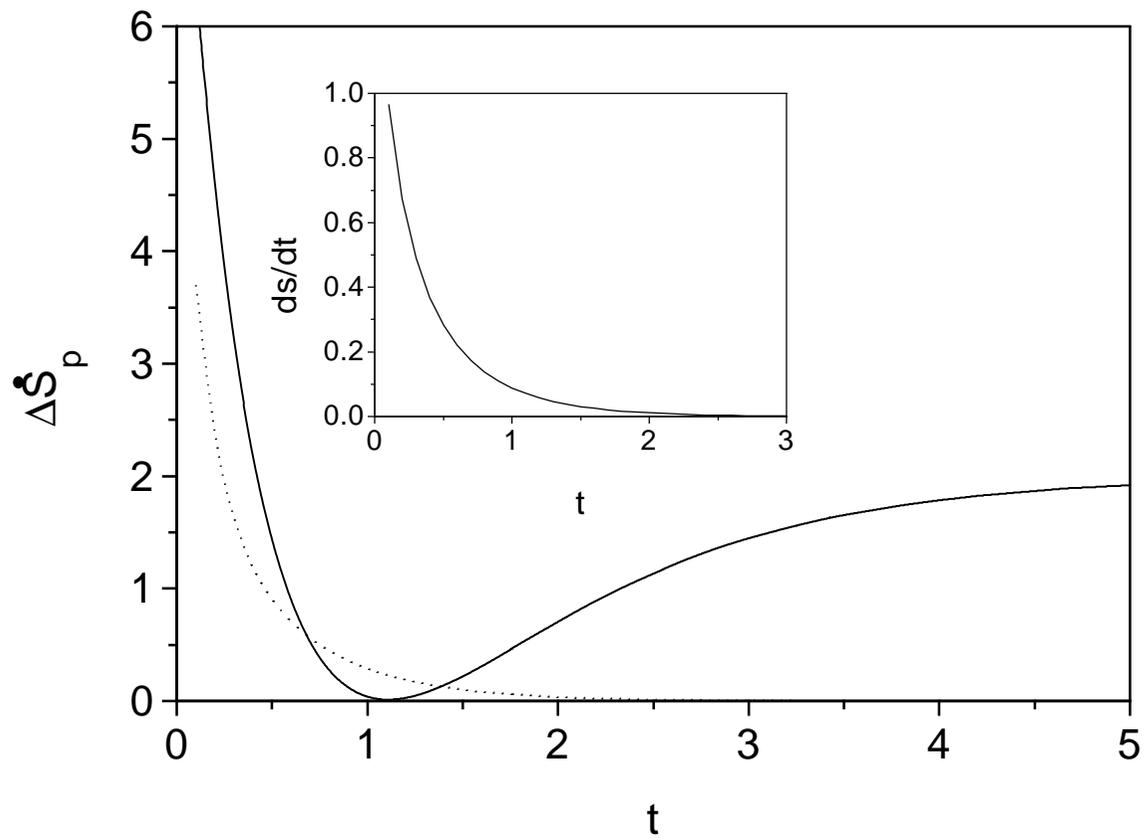
\begin{figure}
\caption{
Plot of entropy production ($\dot{\Delta S_P}$) vs time using Eq.(49) for
the same parameter set as in Fig.1 and $\beta_h(0) =1.0$ and $f_e =1.0$.
$\tau=2$ and 1 for solid and dotted curves. In the inset the sum of 
$\dot{\Delta S_P}$ and $\dot{\Delta S_F}$ from Eqs. (49) and (48) is
plotted against time for $\tau=2$ (Units are arbitrary).
}
\end{figure}

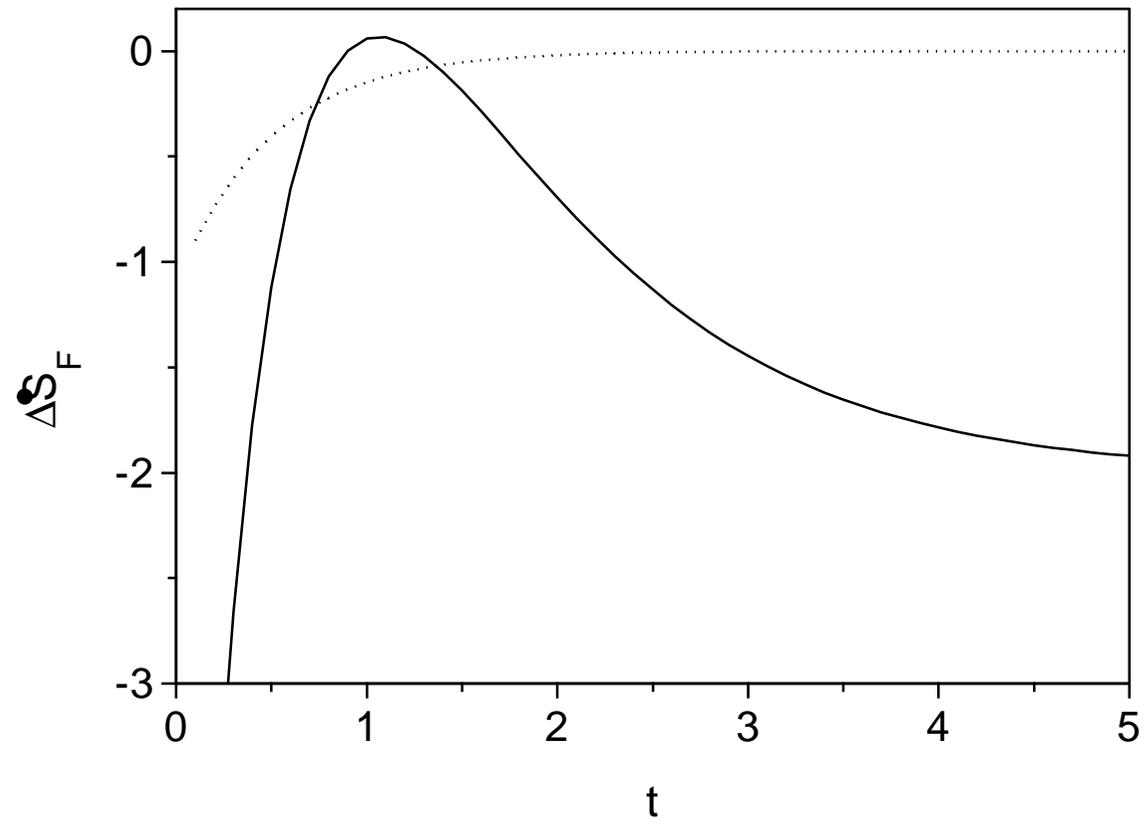
\begin{figure}
\caption{
Plot of entropy flux ($\dot{\Delta S_F}$) vs time using Eq.(48) for
the same parameter set as in Fig.2 
$\tau=2$ and 1 for solid and dotted curves (Units are arbitrary).
}
\end{figure}

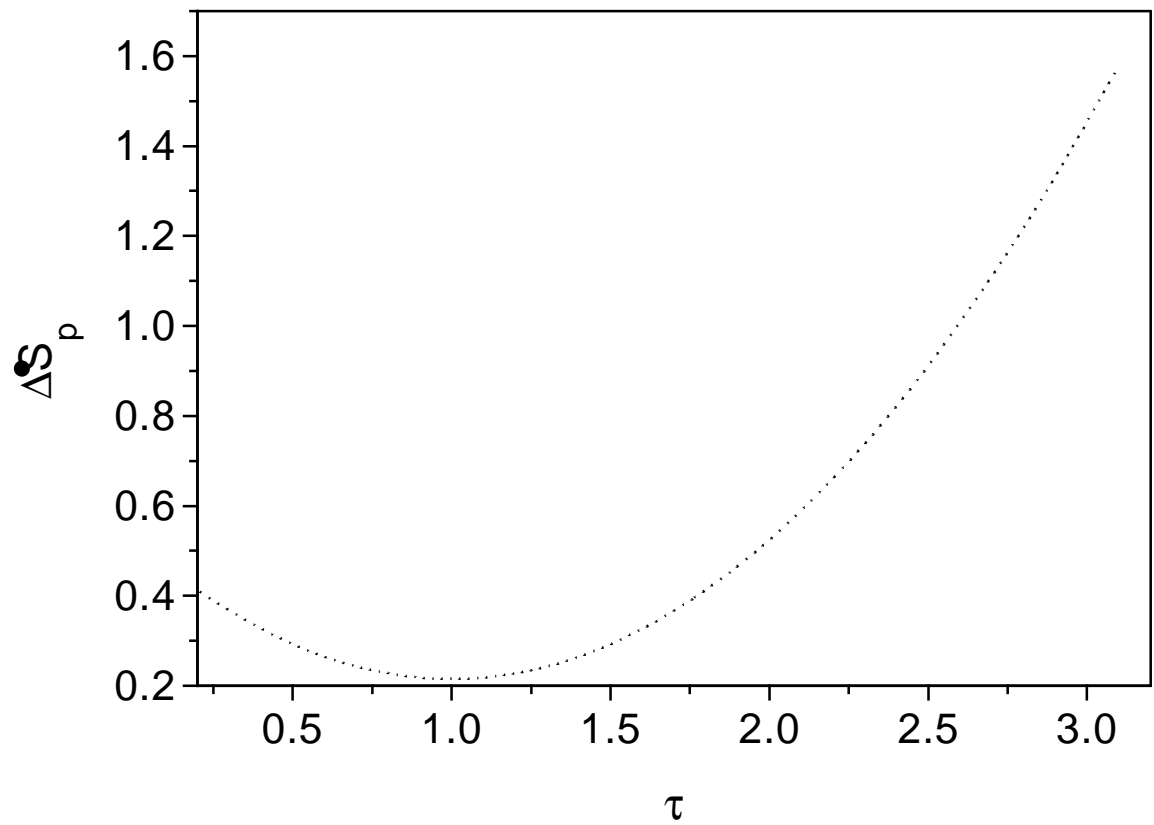
\begin{figure}
\caption{
Plot of entropy Production ($\dot{\Delta S_P}$) vs $\tau$ using Eq.(49) for
the same parameter set as in Fig.2 at $t=0.5$ (Units are arbitrary).
}
\end{figure}

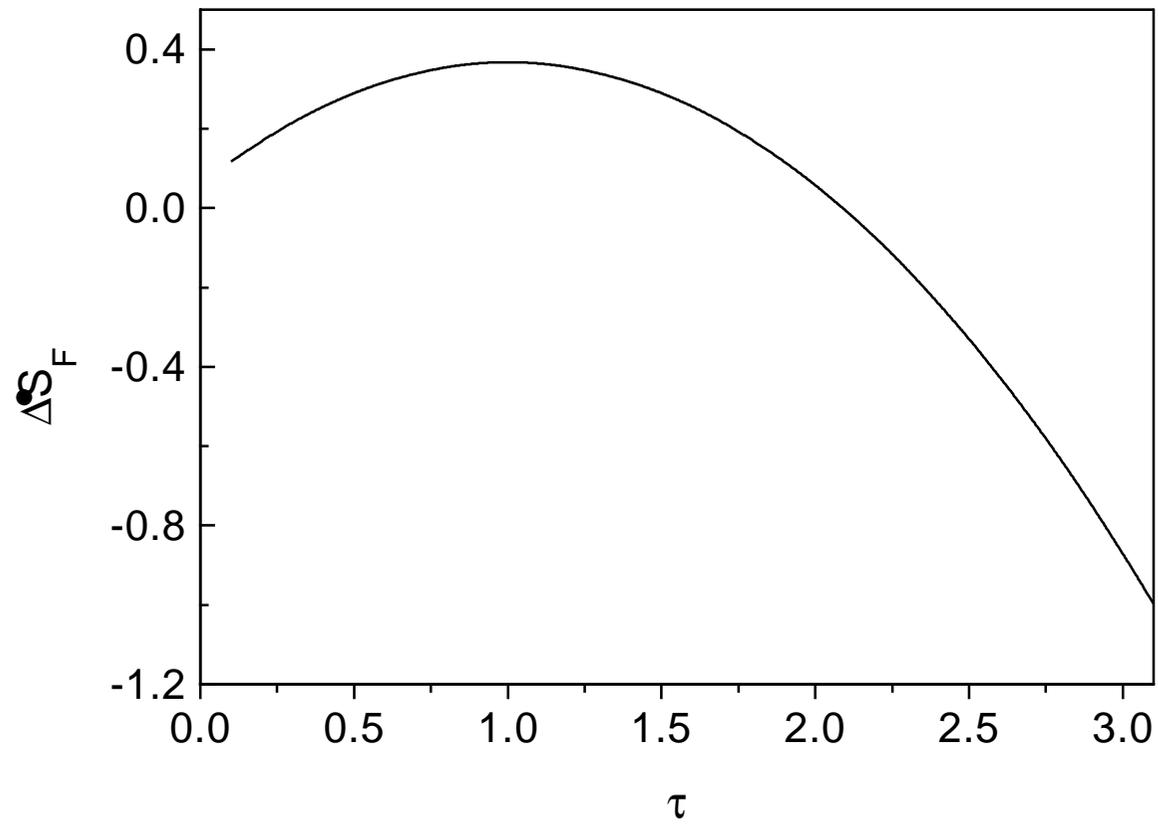
\begin{figure}
\caption{
Plot of entropy flux ($\dot{\Delta S_F}$) vs $\tau$ using Eq.(48) for
the same parameter set as in Fig.2 at $t=0.5$ (Units are arbitrary).
}
\end{figure}

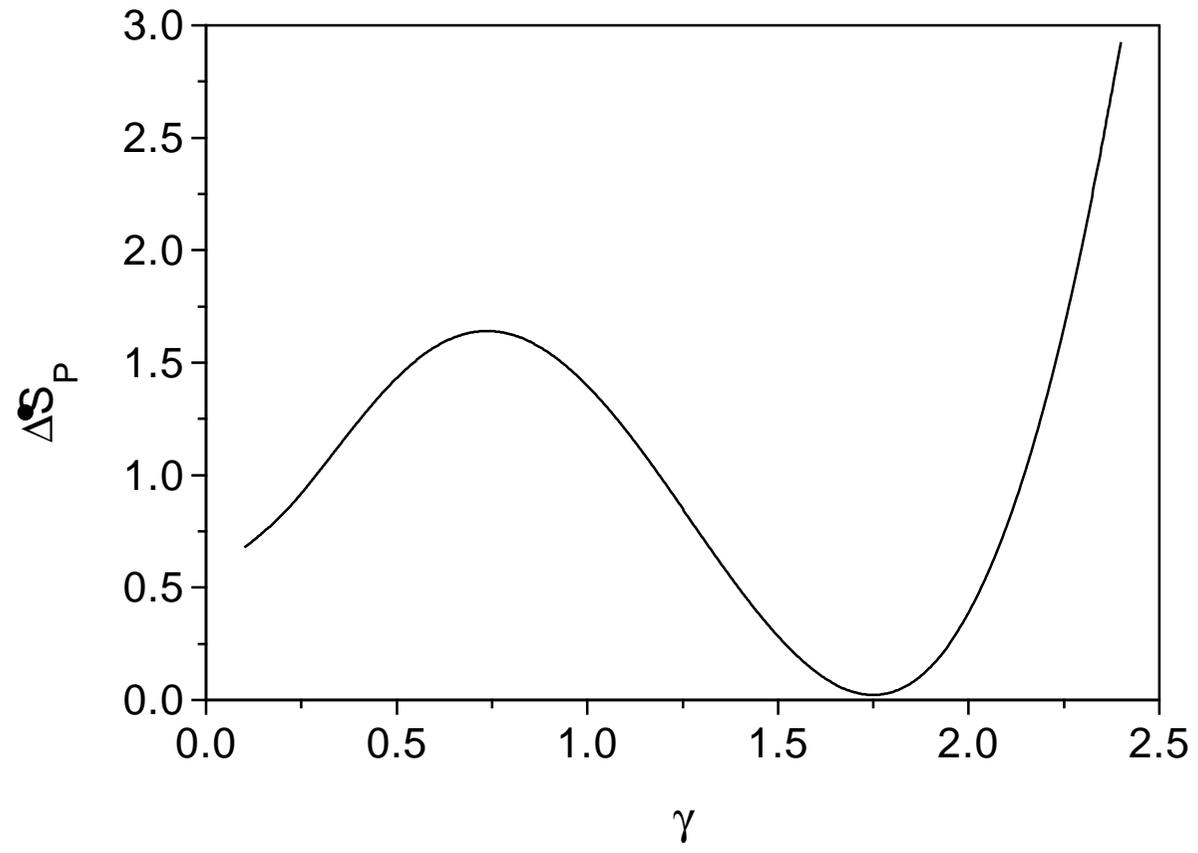
\begin{figure}
\caption{
Plot of entropy Production ($\dot{\Delta S_P}$) vs $\gamma$ using 
Eq.(49) for
the same parameter set as in Fig.2 and $\tau =2$ at $t=0.5$
(Units are arbitrary).
}
\end{figure}

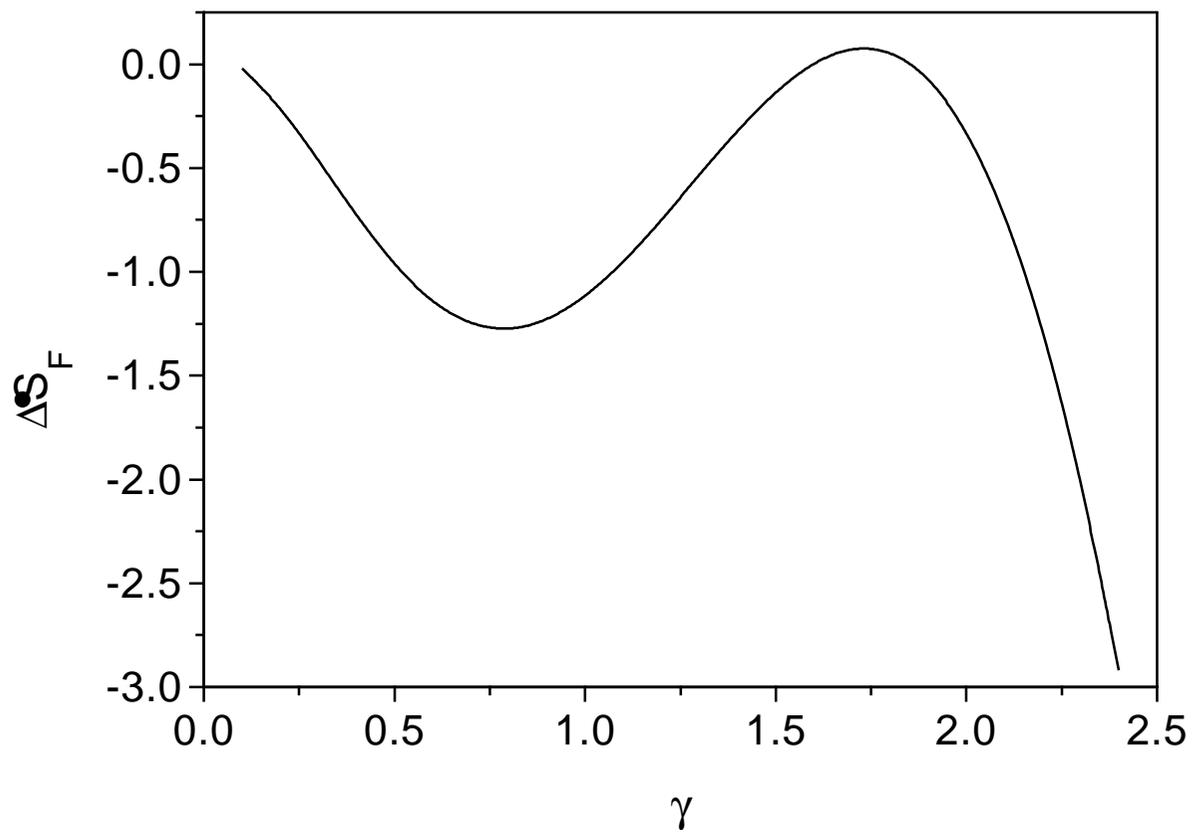
\begin{figure}
\caption{
Plot of entropy flux ($\dot{\Delta S_F}$) vs $\tau$ using Eq.(48) for
the same parameter set as in Fig.6 at $t=0.5$ (Units are arbitrary).
}
\end{figure}

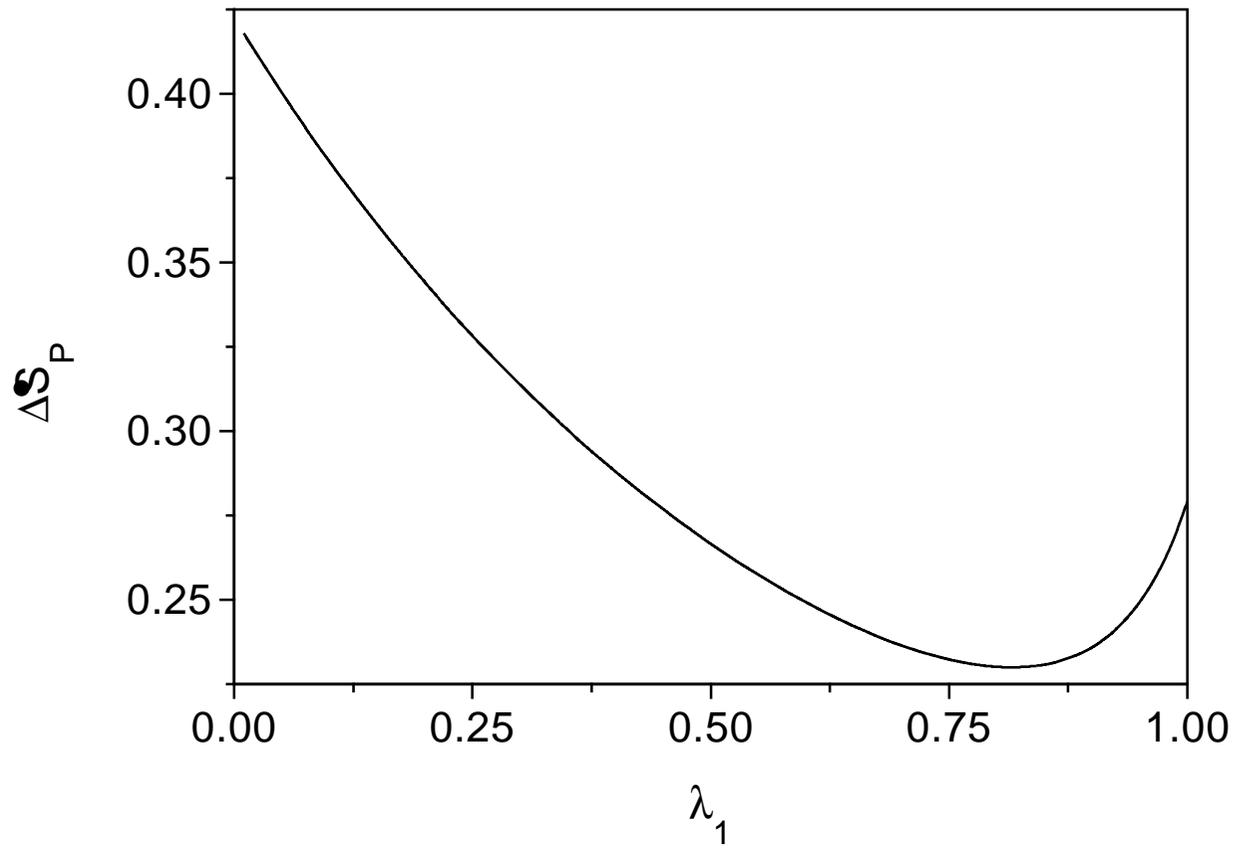
\begin{figure}
\caption{
Plot of entropy Production ($\dot{\Delta S_P}$) vs $\lambda_1$ using Eq.(59) 
for $\sigma_1=0.0$, $\beta_h'=0.0$, $D'=1.0$, $\alpha=1.0$ and $\gamma=1.0$ 
at $t=0.5$ (Units are arbitrary).
}
\end{figure}

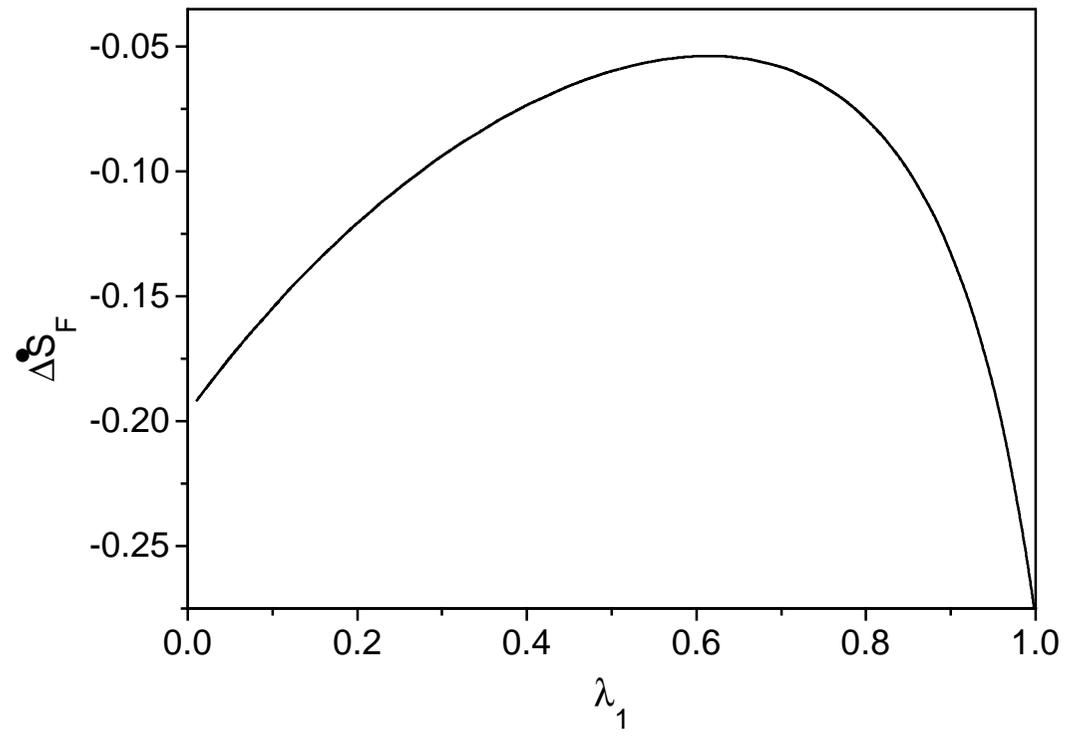
\begin{figure}
\caption{
Plot of entropy flux ($\dot{\Delta S_F}$) vs $\lambda_1$ using Eq.(58) for
the same parameter set as in Fig.8 at $t=0.5$ (Units are arbitrary).
}
\end{figure}


\end{document}